\documentclass{elsart1p}
\usepackage{graphicx}
\usepackage{amssymb,amsmath,amsopn,bm,bbm}

\DeclareMathOperator{\im}{Im}

\newcommand{\dd}{\mathrm{d}}

\newcommand{\erw}[1]{\left \langle #1 \right \rangle}
\newcommand{\GeV}{\mathrm{GeV}}
\newcommand{\MeV}{\mathrm{MeV}}

\begin{document}
\begin{frontmatter}
%
%
%
\title{Dileptons in Heavy-Ion Collisions}
%

\author[hvh]{Hendrik van Hees}
\author[rr]{Ralf Rapp}

\address[hvh]{Institut f{\"u}r Theoretische Physik,
  Justus-Liebig-Universit{\"a}t Giessen, D-35392 Giessen, Germany}
\address[rr]{Cyclotron Institute and Physics Department, 
Texas A\&M University, College Station, Texas 77843-3366, U.S.A.}
%
%
%
\begin{abstract}
  Due to their penetrating nature, dileptons are a valuable probe for
  the properties of the hot and/or dense medium created in relativistic
  heavy-ion collisions. Dilepton invariant-mass spectra provide
  direct access to the properties of the electromagnetic
  current-correlation function in strongly interacting matter. In this
  paper an overview is given of our current theoretical understanding of
  the dilepton phenomenology in comparison to recent data in heavy-ion
  collisions at the CERN SPS.  
\end{abstract}
\end{frontmatter}
%
\section{Introduction}
\label{sec.intro}

Electromagnetic (EM) probes, i.e., photons and lepton pairs (dileptons),
do not participate in the strong interaction and can therefore mediate
valuable information on the EM current correlator in the interior of the
hot and dense fireball created in ultrarelativistic heavy-ion collisions
(URHICs)~\cite{Rapp:1999ej}; their spectra are nearly unaffected by
final-state interactions.

In this paper our present theoretical understanding of recent data on
invariant-mass ($M$) and transverse-momentum ($q_t$) spectra of
dileptons in URHICs at the CERN SPS and BNL RHIC will be reviewed. The
emission rate of dileptons with an invariant mass $M=(q_0^2-q^2)^{1/2}$
and three-momentum, $q$, from a medium at temperature $T$ is given
by\cite{Shuryak:1980tp,McLerran:1984ay,Gale:1990pn}
\begin{equation}
\label{MT}
\frac{\dd N_{ll}}{\dd^4 x \dd^4 q} = -\frac{\alpha^2}{3 \pi^3}
\frac{L(M)}{M^2} \im \Pi_{\mathrm{em}\mu}^{\mu}(M,q;T,\mu_B) f_B(q_0,T),
\end{equation}
where $\alpha \simeq 1/137$ denotes the fine-structure constant, $L$ the
lepton-phase space factor, $\Pi_{\mathrm{em}}^{\mu \nu}$ the retarded
in-medium EM current correlator, and $f_B$ the Bose-Einstein
distribution. In the vacuum the hadronic EM correlator can be inferred
from $\mathrm{e}^+ \mathrm{e}^- \rightarrow \mathrm{hadrons}$. At low
$M$ it is well-described by the vector dominance model (VDM),
including the light vector mesons, $\rho$, $\omega$, and $\phi$ and at
higher $M \gtrsim 1.5 \;\GeV$ by the perturbative QCD (pQCD) continuum.

The theoretical investigation of the dilepton signal in URHICs thus must
aim at a concise model for the spectral properties of the light vector
mesons, most importantly the $\rho$ meson in the iso-vector-vector
channel, which give the most important contribution to the EM current
correlator, and of the Quark-Gluon Plasma (QGP). The remainder of this
paper is organized as follows: In Sec.~\ref{sec_medium} theoretical
models for the EM current correlator in partonic as well as hot and/or
dense hadronic matter are summarized, followed by a comparison 
to recent dilepton data from the SPS 
in Sec.~\ref{sec_dileps-hics} and conclusions and outlook in
Sec.~\ref{sec_concl}.

\section{In-medium properties of the EM current correlator}
\label{sec_medium}
Approximate chiral symmetry (CS) in the light-quark sector of QCD is one
of the most important ingredients for the building of effective hadronic
models. In the vacuum and at low temperatures and/or densities CS is
broken dynamically through the formation of a quark condensate,
$\erw{\bar{\psi} \psi} \neq 0$, leading to the mass splitting of the
mass spectra of chiral hadron multiplets. One of the most evident
manifestations of CS breaking is seen in the measurement of the
isovector-vector and -axialvector current correlators through $\tau
\rightarrow \nu + n \pi$-decay data ($n$: number of 
pions)~\cite{Barate:1998uf,Ackerstaff:1998yj}. Finite-temperature
lattice-QCD (lQCD) calculations~\cite{Fodor:2004nz,Karsch:2007vw} find a
melting of the quark condensate with increasing temperature and the
restoration of CS above a critical temperature, $T_c \simeq
160$-$190\;\MeV$. Another finding is that the CS restoration (CSR) and
deconfinement transition temperatures,
coincide~\cite{Karsch:1994hm}. From these findings one expects significant
changes of the hadron spectra in a hot and dense medium close to CSR.

In the literature, two scenarios concerning the manifestation of CSR in
the hadron spectrum have emerged: In one scenario it has been suggested
that due to the melting of the quark condensate hadron masses should
drop to $0$ at the critical point~\cite{Brown:1991kk}. The other
mechanism is found in phenomenological hadronic many-body models, where
hadron spectral functions show a significant broadening with small mass
shifts~\cite{Gale:1990pn,Gale:1993zj,Rapp:1997fs,Rapp:1999us}. It turns
out, however, that both scenarios, i.e., either dropping masses or the
broadening of in-medium hadron widths (``melting resonances'') are
compatible with QCD-sum rule
calculations~\cite{Asakawa:1993pq,Leupold:1997dg,Klingl:1997kf,Ruppert:2005id}. Thus
the dilepton signal in URHICs gives an important experimental insight
into the nature of the CSR through the vector part of the EM current
correlator. Since on the other hand a direct measurement of the
axialvector correlation function in heavy-ion collisions is difficult, a
direct assessment of CSR in HIC's seems not to be possible. Thus, a
promising theoretical evidence for CSR may be achievable by the
application of finite-temperature Weinberg-sum
rules~\cite{Weinberg:1967kj,Kapusta:1993hq} which relate moments of the
difference of vector and axial-vector spectral functions to order
parameters of CS like quark and four-quark condensates, which are in
principle accessible in lQCD simulations.

A model-independent approach to the EM current correlator is the use of
the chiral-reduction formalism based on a low-density (virial)
expansion~\cite{Steele:1996su} to evaluate medium modifications with
empirical vacuum-vector and axialvector-correlation functions as
input. In this approach a mechanism for CSR is the mixing of the vector
and axialvector correlators through pions in the medium, similar to the
``chiral mixing'' found on the basis of current algebra and
PCAC~\cite{Dey:1990ba}. However, the applicability of such methods is
limited to the low-density/temperature region of the medium. Thus the
building of effective hadronic models and the application of
quantum-field theory methods becomes necessary to assess the dilepton
signal in URHICs realistically. The most important guideline for
building effective hadronic models is CS, e.g., using (generalized)
hidden local symmetry~\cite{Bando:1984ej,Harada:2003jx} to describe
vector (and axial-vector) mesons as gauge bosons. Recently, it has been
shown that within these models CS can be realized in the vector
manifestation (VM), leading to a field-theoretical model for the
dropping-mass scenario of CSR~\cite{Harada:2005br}. However, the same
model also admits the usual Wigner-Weyl realization, where one finds
degeneracy of the $\rho$ and $a_1$ spectral functions with little mass
shifts~\cite{Sasaki:2009ey}. In general, this finding suggests that a
decision in which way CS is realized in nature and how it is restored
cannot be achieved from the fundamental principle of CS alone.

Another approach is the use of phenomenological Lagrangians which
describe the vacuum properties of the vector mesons and evaluate medium
modifications of their spectral functions within
finite-temperature/density quantum field theory. To account for the
strong couplings, hadronic many-body theory (HMBT) implements
non-perturbative techniques like the dressing of, e.g., the pion
propagator in the $\rho$-meson's pion cloud, as well as a resummation of
direct interactions of the $\rho$ with mesons and baryons of the medium
(for a review see~\cite{Rapp:1999ej}). One finds small mass shifts of
the vector mesons due to many repulsive and attractive interactions with
cancellations in the real part of the vector-meson self-energy, but a
substantial broadening of their spectral functions. An interesting
result of the model in Ref.~\cite{Rapp:1999us} is the apparent
degeneracy of the pertinent hadronic dilepton-emission rates and that of
hard-thermal-loop improved pQCD rates~\cite{Braaten:1990wp} at
temperatures close to the critical region, $T_c \simeq 160$-$190 \;
\MeV$, in a kind of ``quark-hadron duality'', implying CSR through
``resonance melting''. This finding is consistent with the smoothness of
the isovector quark-number susceptibility in lQCD
simulations~\cite{Allton:2005gk}.

Another possibility to assess in-medium properties of vector mesons 
is to employ empirical scattering amplitudes and dispersion-integral 
techniques within the $T\varrho$ approximation for the in-medium
selfenergies~\cite{Eletsky:2001bb}.

\section{The dilepton signal in heavy-ion collisions}
\label{sec_dileps-hics}
To confront the in-medium EM spectral functions from the above 
models to dilepton $M$ and $q_t$ spectra in URHICs a description of the
entire evolution of the produced medium in its hadronic and partonic
stages is necessary. The success of (ideal) hydrodynamics in the
evaluation of the bulk of this matter implies local thermal equilibrium,
i.e., the medium can be modeled by an energy-density and collective-flow
field. In~\cite{vanHees:2006ng,vanHees:2007th} a thermal fireball
parameterization has been used. After a formation time the hot and dense
matter is described as an ideal gas of quarks and gluons, evolving
through a mixed phase to a hadron-resonance gas at a transition
temperature of $T_c \simeq 160$-$190\;\MeV$. As thermal-model
evaluations of particle abundances in URHICs indicate, the chemical
freeze-out temperature is about $T_{\mathrm{ch}} \simeq
160$-$175\;\MeV$~\cite{Andronic:2005yp,Becattini:2005xt}, below which
the particle ratios are fixed through the introduction of chemical
potentials. The thermal freeze-out temperature, around which also elastic
rescatterings cease, occurs at temperatures of $T_{\mathrm{fo}} \simeq
90$-$130 \; \MeV$. The evolution of the medium is parameterized as a
cylindrical homogeneous fireball which is longitudinally and radially
expanding. The temperature is given via the assumption of isentropic
expansion and the equation of state of an ideal gas of massless gluons
and $N_f=2.3$ effective quark flavors with the total entropy fixed by
the number of charged particles. After a mixed phase, which is described
by a standard volume partition, the hadronic phase is modeled as a
hadron-resonance gas.

Early on, measurements of dilepton $M$ spectra in URHICs at the SPS
have shown an indication of medium modifications via an increased yield
in the low-mass region (LMR), $M \leq 1\;\GeV$, compared to expectations
from $pp$ collisions~\cite{Adamova:2002kf}. However, only the recent
precision achieved in the dimuon measurement of the NA60 collaboration
in $158 \, A\GeV$ In-In
collisions~\cite{Damjanovic:2007qm,Arnaldi:2007ru} has made it possible
to subtract the ``hadron-decay cocktail'' contribution to obtain the excess
spectrum and thus to significantly discriminate between models of  
the dropping-mass or the resonance-melting scenarios for CSR. The data
clearly favor small mass shifts and a substantial broadening
of the vector-meson spectral functions.
\begin{figure}
\includegraphics[height=5.0cm]{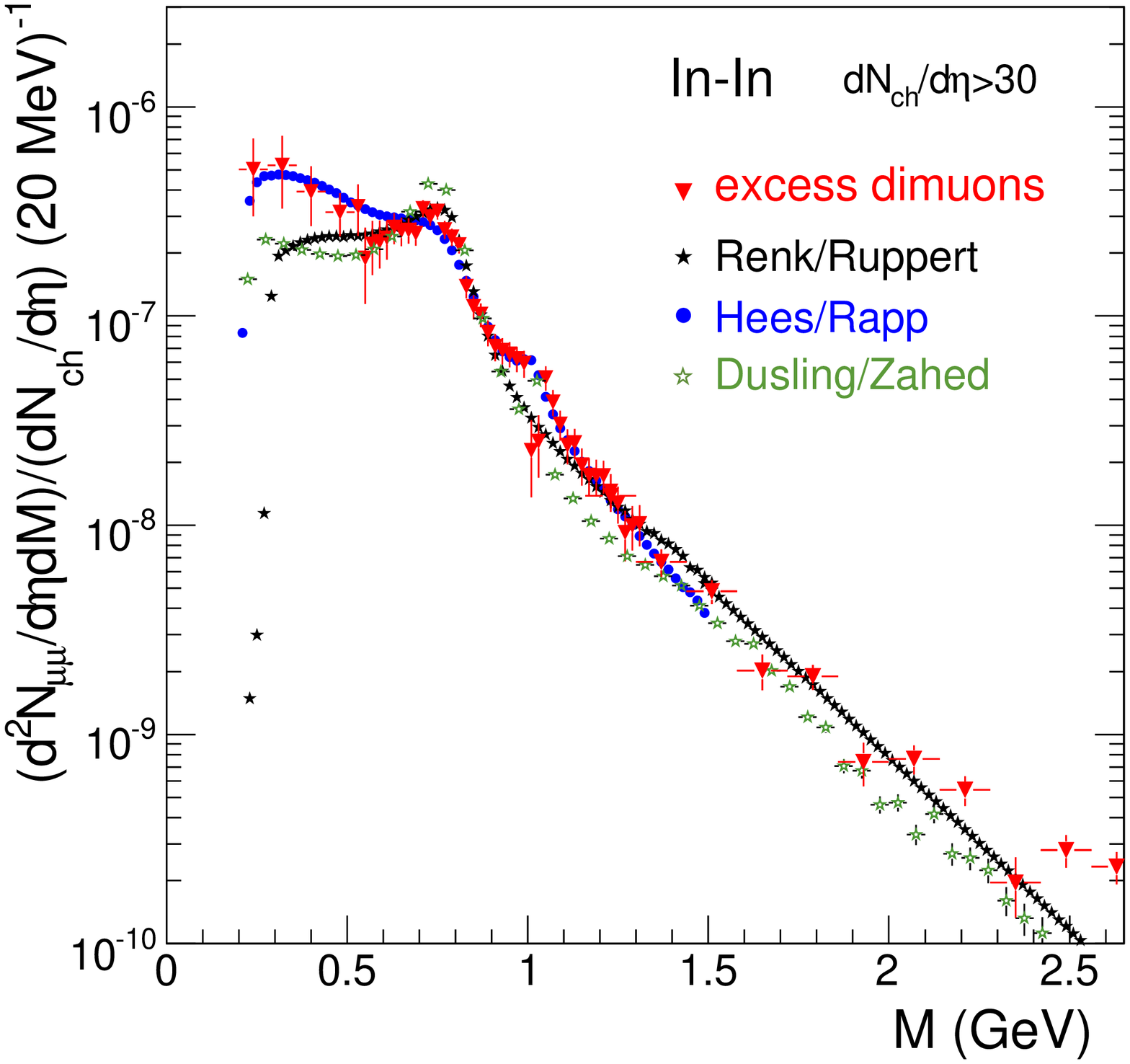} \hspace*{5mm}
\includegraphics[height=5.0cm]{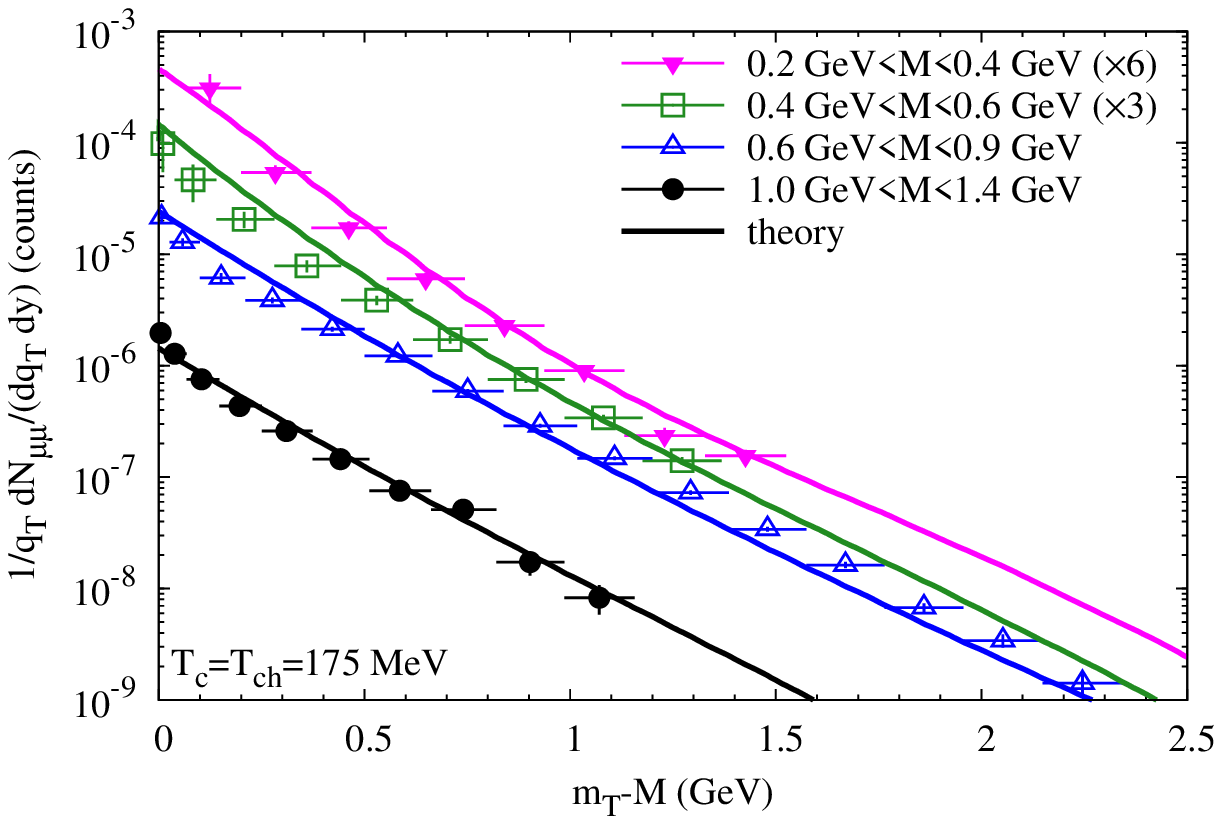}
\caption{(Color online) Left panel: Dimuon excess $M$ spectrum in $158
  \, A\GeV$ In-In collisions~\cite{Arnaldi:2008er} compared to model
  calculations
  in~\cite{vanHees:2007th,Ruppert:2007cr,Dusling:2007su}. Right panel:
  $q_t$ spectra~\cite{Arnaldi:2008fw} in various $M$ bins compared to
  the model in~\cite{vanHees:2007th} (using EoS-A and a radial
  acceleration $a_{\perp}=0.085\;c^2/\mathrm{fm}$ in the fireball). 
  Note that the measurements refer to centrality-inclusive
  data, while the calculation assumes semi-central collisions
  defined by $\dd N_{\mathrm{ch}}/\dd y = 140$.}
\label{fig.1}
\end{figure}
In Fig.~\ref{fig.1} we compare the absolutely normalized mass
spectrum~\cite{Arnaldi:2008er} to the following models (i)
Ref.~\cite{vanHees:2007th}, using spectral functions from
HMBT~\cite{Rapp:1999us} supplemented with thermal radiation from
multi-pion induced processes and non-thermal sources (Drell-Yan
annihilation, decays of primordial $\rho$ mesons not equilibrated with
the medium and $\rho$ decays after thermal freeze-out), (ii)
Ref.~\cite{Ruppert:2007cr}, implementing spectral functions from the
$T\varrho$ approach in~\cite{Eletsky:2001bb}, and (iii)
Ref.~\cite{Dusling:2007su}, where the chiral-reduction formalism has
been applied. For the medium evolution in (i) and (ii) fireball models
are used, while in (iii) a hydrodynamical calculation is employed. The
predicted broadening of the EM correlator, particularly in model (i) at
low masses, is dominated by baryonic excitations of the vector mesons in
the medium.

In the same figure, $m_t$ spectra ($m_t^2=M^2+p_t^2$) in different mass
bins have been compared to the model in~\cite{vanHees:2007th}. The
latter also contains studies of the sensitivity to uncertainties in (a)
$T_c$ as determined in present lQCD simulations, and (b)
$T_{\mathrm{ch}}$ as extracted from thermal models for hadron
production.  Three equations of state with $T_c=T_{\mathrm{ch}}=175 \;
\MeV$ (EoS-A), $T_c=T_{\mathrm{ch}}=160\;\MeV$ (EoS-B), and $T_c=190
\;\MeV$, $T_{\mathrm{ch}}=160 \; \MeV$ (EoS-C) have been used. In EoS-C
a chemically equilibrated hadronic phase in the temperature region
$160\;\MeV \leq T \leq 190 \; \MeV$ has been assumed. The comparison to
the $M$ and $m_t$ spectra is qualitatively comparable to that of EoS-A
(up to small variations of the total dilepton yields which could be
readjusted by small changes of the fireball lifetime). An interesting
consequence of this insensitivity to $T_c$ and $T_{\mathrm{ch}}$ is that
the dimuon spectrum in the intermediate-mass region (IMR) $M \geq
1\;\GeV$ can be equally well described with models where the dilepton
yield is either dominated by radiation from a partonic (EoS-B) or a
hadronic (EoS-C) source. Since the emission in this $M$ region is
dominated from fireball stages with temperatures around the critical
region, $T \simeq 160$-$190 \; \MeV$, this insensitivity is due to the
above described ``parton-hadron duality'' of the dilepton rates within
this model. Thus, a definite conclusion whether the dileptons in the IMR
are dominated by radiation from a partonic or hadronic medium can only
be drawn if a more precise value of $T_c$ is known.

\begin{figure}
\includegraphics[height=4.5cm]{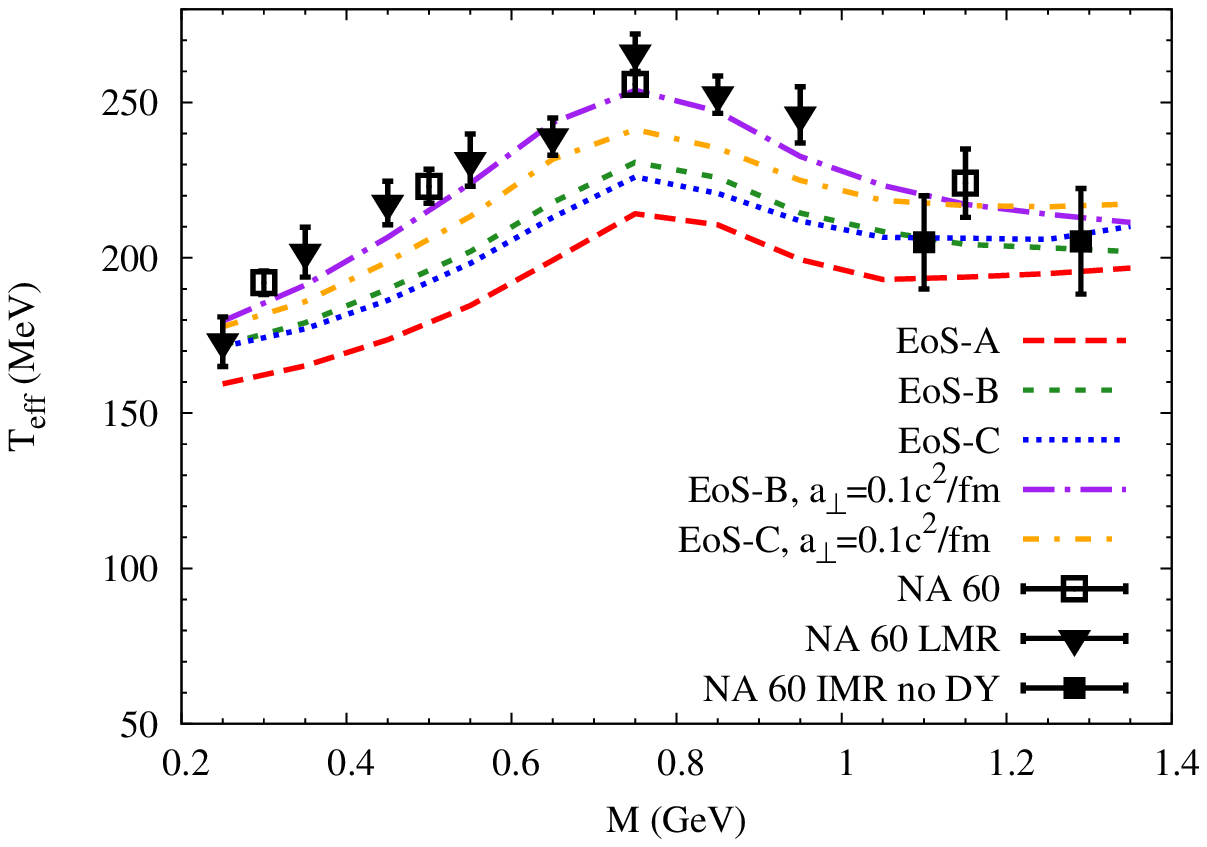} \hspace*{5mm}
\includegraphics[height=4.5cm]{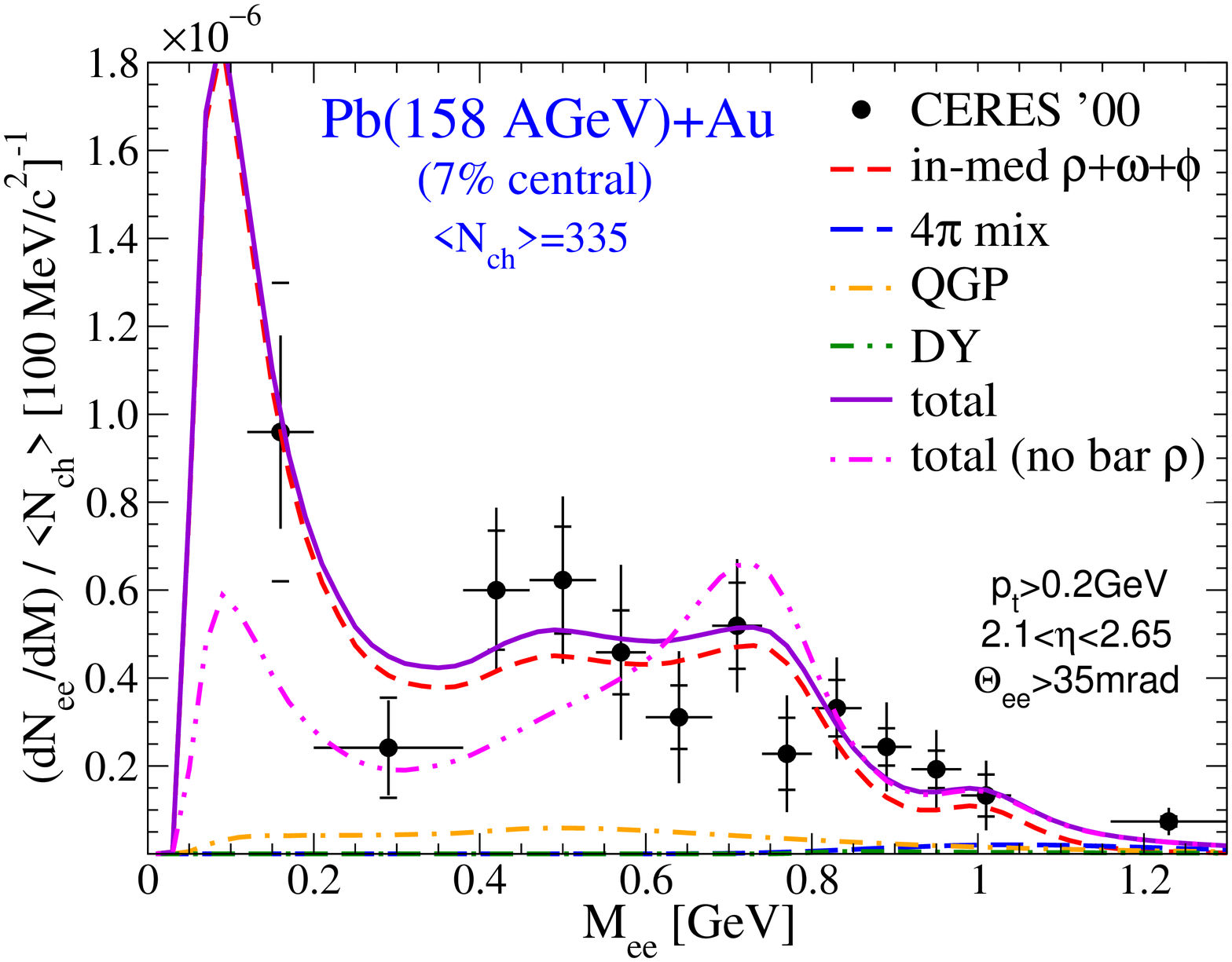}
\caption{(Color online) Left panel: effective slopes from $m_t$ spectra
  for different equations of state and radial acceleration of the
  fireball~\cite{vanHees:2007th}, compared to NA60 
  data~\cite{Damjanovic:2007qm,Arnaldi:2007ru}. 
  Right panel: Comparison of the same model~\cite{vanHees:2007th} to 
  recent CERES/NA45 dielectron spectra~\cite{Adamova:2006nu} in central
  $158 A\GeV$ Pb-Au collisions.}
\vspace*{-3mm}
\label{fig.2}
\end{figure}

As shown in the left panel of Fig.~\ref{fig.2} the comparison of the
effective slopes of the $m_t$ spectra with the corresponding analysis of
the NA60 data indicates that a larger radial flow of the medium is
favored by the data.
The right panel of Fig.~\ref{fig.2} shows a comparison of the same model
to the recent dielectron spectrum in central $158 A\GeV$ Pb-Au
collisions from the NA45/CERES collaboration~\cite{Adamova:2006nu}.  The
result of the same model \emph{without} the interactions of the $\rho$
meson with baryons in the medium corroborates their prevalence for the
broadening of the $\rho$-spectral function, in a very pronounced way in
the mass region below the two-pion threshold.

\section{Conclusions and Outlook}
\label{sec_concl}
The comparison of effective hadronic models for in-medium properties of
the EM current-correlation function with high precision dilepton data in
URHICs is a promising method to gain insights in the nature of
CSR. Models based on hadronic many-body theory, predicting a strong
broadening of the vector-meson spectral functions with little mass
shifts, are favored by recent measurements compared to those
implementing the dropping-mass conjecture. However, a more complete
analysis of the generalized hidden-local symmetry model with the vector
manifestation of CS, leading to dropping vector and axialvector masses,
including baryonic interactions is not available yet. The large
enhancement of the dilepton yield in the LMR, recently observed by
PHENIX in $200 A\GeV$ Au-Au collisions at RHIC~\cite{Afanasiev:2007xw},
to date cannot be explained by any of the models which are successful at
the SPS.

The extension of the present models to axialvector mesons, constrained
by lQCD calculations of chiral order parameters in connection with
Weinberg sum rules, might help to deduce more direct evidence for CSR
from the dilepton signal in URHICs.

\textbf{Acknowledgments.} One of us (HvH) likes to thank the organizers
for the invitation and generous support to attend an exciting
conference. This work was supported in part by the U.S. National Science
Foundation under grant no. PHY-0449489.


\begin{flushleft}

\end{flushleft}

%
%
%
%
%
%
%
%
%
\end{document}